\documentclass[twocolumn,showpacs,preprintnumbers,amsmath,amssymb]{revtex4}
\usepackage[xdvi]{graphicx}%
\usepackage{dcolumn}
\usepackage{amsmath}

\newcommand{\bra}{\left\langle}
\newcommand{\ket}{\right\rangle}

\newcommand{\lr}[1]{\left(#1\right)}

\newcommand{\e}{{\rm e}}

\newcommand{\ep}{\epsilon}

\newcommand{\Dt}{\Delta t}

\begin{document}

\title{Scale-free patterns at a saddle-node bifurcation
in a stochastic system}

\author{Mami Iwata}
\email{iwata@jiro.c.u-tokyo.ac.jp}
\author{Shin-ichi Sasa}
\email{sasa@jiro.c.u-tokyo.ac.jp}
\affiliation
{Department of Pure and Applied Sciences,
University of Tokyo, 3-8-1 Komaba Meguro-ku, Tokyo 153-8902, Japan}
\date{Received 6 June 2008; published 14 November 2008}


\begin{abstract}
We demonstrate that scale-free patterns are observed in a spatially 
extended stochastic system whose deterministic part undergoes 
a saddle-node bifurcation. Remarkably, the scale-free patterns appear only 
at  a particular time in relaxation processes from a spatially homogeneous 
initial condition. We characterize the scale-free nature in terms of 
the spatial configuration of the exiting time from a marginal saddle 
where the pair annihilation of a saddle and a node occurs at the 
bifurcation point. Critical exponents associated with the scale-free 
patterns are determined by numerical experiments.
\end{abstract}

\maketitle



Scale-free patterns are widely observed in nature.
The most familiar one in statistical mechanics 
might be that in particle configurations at the gas-liquid 
critical point \cite{Stanley}.
The patterns are characterized by a correlation
length which is given by a power-law function of 
the distance from the critical temperature. 
Another type of scale-free patterns was found
in diffusion-limited aggregation 
\cite{Herrmann} and spinodal decomposition \cite{Bray}.
In this type, the length scale that characterizes 
the patterns grows  as a power-law function of time, without 
the fine-tuning of the system parameters.


Recently, a new type of scale-free pattern  
called {\it dynamical heterogeneity} has been found in glassy systems
\cite{ryamamoto, BB, berthier, garrahan, shear, colloid_sim, MCT}.
Such patterns are fascinating  because they become 
visible only by quantifying some dynamical events
(called {\it bond-breaking events} or {\it unlocking events})
during a particular time interval $t_*$. 
Let us consider dense colloidal suspensions as an example.
We denote the displacement of particles
in a region around a position $x$ during a time interval $t$
as $q(x,t)$.
Then, the correlation length $\xi(t)$ of the pattern $q(x,t)$ 
becomes maximum at a time $t_*$ and $\xi(t_*)$ exhibits 
 divergent behavior as a function of the distance from 
a critical parameter value.


The characterization in terms of the time-dependent 
correlation length $\xi(t)$ is similar to that of 
growing patterns, while the existence of the critical 
parameter value 
is similar to 
critical phenomena in equilibrium systems. 
Such a coexistence of these features implies that
$q(x,t_*)$ indeed belongs to a new type of scale-free patterns.
Since its nature is simple and nontrivial, we  expect 
that there exists a wide class of systems that exhibit 
such a type of patterns. Motivated by this expectation, 
we  propose the simplest model among them. We then intend 
to elucidate the nature of the new type of scale-free 
patterns.

\paragraph*{Model:}


In the present paper, we study a spatially extended stochastic system 
that exhibits relaxation behavior at a saddle-node bifurcation.
Although there is a certain idea  that connects this model with 
the understanding of glassy systems \cite{jstat,epl}, we do not 
consider the relationship in this paper. Instead, we regard this 
model as a typical system that undergoes an elementary bifurcation 
under the influence of noise. By elementary bifurcation,
we imply  pitch-fork bifurcation,  Hopf bifurcation, and  
saddle-node bifurcation \cite{Gucken}. 
Among them, the first two bifurcations under the 
influence of noise have been studied intensively  
in the the context of critical phenomena 
\cite{Nigel,Kuramoto,Daido}.
Therefore, a saddle-node bifurcation  with noise 
might be related to a new class of critical phenomena.

\begin{figure}[htbp]
\includegraphics[width=6cm]{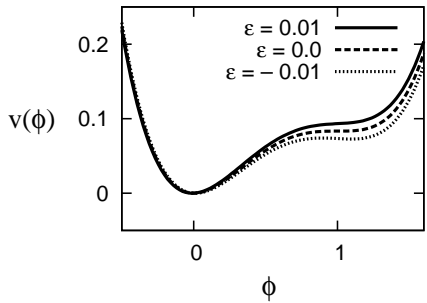}
\caption{
 Potential $v$ as a function of $\phi$ for
several values of $\epsilon$.}
\label{pot_phi}
\end{figure}

Specifically, we study  a coupled Langevin equation in a
one-dimensional lattice $\{i|i=1,2,\cdots, N \} $. 
Let $\phi_i$ be a one-component quantity defined at the $i$-th
site, where we assume periodic boundary conditions $\phi_0=\phi_N$ and 
$\phi_{N+1}=\phi_1$.
Then, $\phi_i$ obeys 
\begin{eqnarray}
\partial_t \phi_i&=& f(\phi_i)
+\kappa \lr{\phi_{i+1}+\phi_{i-1}-2\phi_i}+\xi_i, 
\label{langevin} 
\end{eqnarray}
where 
$
f(\phi)=-\phi((\phi-1)^2+\epsilon)
$
with small $\epsilon$; $\kappa$ is a coupling constant 
and $\xi_i$
represents Gaussian white noise that satisfies
$
\bra\xi_i(t)\xi_j(t')\ket=2T\delta(t-t')\delta_{ij}.
$
The noise intensity $T$ is assumed to be a small positive constant. 
The potential function $v(\phi)$, which is defined by  
$f=- \partial_\phi v$, has a single minimum at $\phi=0$
when $\epsilon \geq 0$, while a pair comprising the 
 minimum  and  maximum appears
around $\phi =1$ when $\epsilon$ becomes negative 
(see Fig.~\ref{pot_phi}). The minimum and maximum correspond to 
the node and saddle point, respectively, 
 in the deterministic
equation $\partial_t  \phi=f(\phi)$. Such a qualitative 
change in the trajectories at $\epsilon=0$ in this deterministic equation
is called {\it saddle-node bifurcation}.  
We call the fixed point $\phi=1$ for $\ep=0$ {\it marginal saddle}
because this saddle 
vanishes when $\ep > 0$. 
A characteristic feature of  (\ref{langevin}) is that even small noise 
drastically affects 
the trajectories passing through the marginal saddle.
 For example, when $\epsilon \le 0$, the deterministic 
trajectories starting from $\phi=1.2$ reach the fixed 
point near $\phi=1$, while trajectories under the influence of noise 
escape from this fixed point and finally arrive at the region near 
the globally minimum point $\phi=0$. 
In the argument below, in order to extract a simple relation, we 
focus on the case $\ep=0$ at which the saddle-node bifurcation occurs.

Since the model we study might be the simplest in the class
of spatially extended stochastic systems in which a saddle-node 
bifurcation occurs locally in space, our model has some relevance to many
experimental systems. One example is  fluctuating motion of a single 
one-dimensional polymer that is subjected
to a trapping potential 
in a liquid  (see Ref. \cite{polymer}). Another example might be 
found in stochastic reaction-diffusion systems because a  saddle-node 
bifurcation is observed in
chemical reactions (see Ref. \cite{Tyson} as an illuminating example).  
Although multiplicative noise  generally occurs in chemical reaction systems, 
we expect that the multiplicative nature does not affect the behavior reported 
below unless the noise intensity becomes zero at the marginal saddle. 


We investigate  (\ref{langevin}) by numerical simulations. Concretely, 
we employ an explicit discrete method with a time step $\Dt=0.01$, 
where the error is $O(\Dt^{3/2})$. We have confirmed that the 
important results reported below are  kept to be valid when 
we select $\Dt=0.001$. We also restrict our investigations to 
the case $\kappa=1$ 
 We obtained the same result for the case $\kappa=0.5$.
That is, a fine-tuning of the parameter value is not necessary
\cite{kappa}. 
Finally, since we are interested in the patterns emerging from a homogeneous 
state, we assume that $\phi_i(0)=1.2$.

\paragraph*{Preliminaries:}


\begin{figure}[htbp]
\includegraphics[width=6cm]{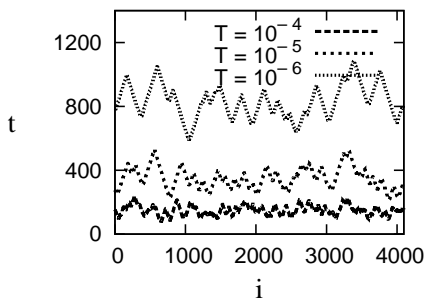}
\caption{Contour curves  defined by $\phi_i(t)=0.5$
for several values of $T$. $N=4096$.
}
\label{phase}
\end{figure} 

\begin{figure}[htbp]
\includegraphics[width=6cm]{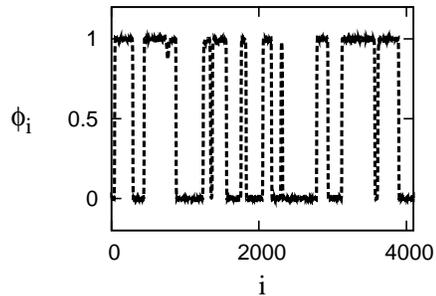}
\caption{Spatial pattern at time $t=828$ 
for the system with $T=10^{-6}$. $N=4096$.
}
\label{domain}
\end{figure} 

In order to observe the time evolution of $\phi_i$, we 
display contour curves defined by $\phi_i(t)=0.5$ for 
several values of $T$ (see Fig.~\ref{phase}). 
Each contour curve distinguishes the late stage ($\phi_i(t) \simeq 0$)
from the early stage ($\phi_i(t) \simeq 1$) because the two 
stages are connected in a short time interval around the 
exiting time from  the marginal saddle at each site $i$.
Thus, for example, 
the pattern $\phi_i$ at time $t=828$ 
for the system with $T=10^{-6}$ 
consists of domains $\phi_i \simeq 1$ and $\phi_i \simeq 0$, 
as shown in Fig.~\ref{domain}, where the time $t=828$
is selected such that the spatial average of 
$\phi_i$ becomes $0.5$.
It should be noted that such a pattern appears only around a particular time 
$t=t_*$. 
Furthermore, the typical length scale of the pattern 
at the time $t_*$ increases as $T$ is decreases. This suggests 
that $\phi_i(t_*)$ is  scale-free in the limit $T \to 0$.



\begin{figure}[htbp]
\includegraphics[width=7cm]{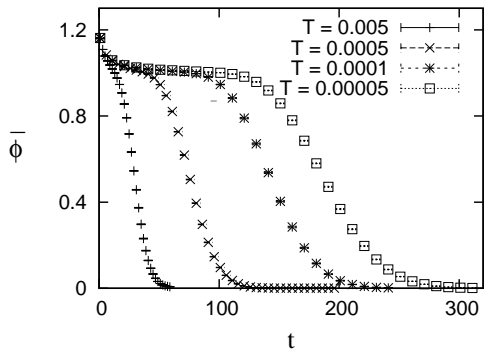}
\caption{ 
$\bar \phi$ as a function of $t$. 
$N=1024$. 
}
\label{phit:fig}
\end{figure}

\begin{figure}[htbp]
\includegraphics[width=7cm]{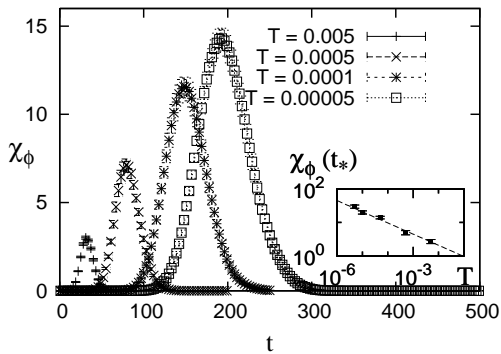}
\caption{
$\chi_{\phi}$ as a function of 
 $t$.
$N=1024$.
Inset: $\chi_{\phi}(t_*)$ as a function of $T$ 
in a log-log scale. The guide line
represents $\chi_{\phi}(t_*)\simeq T^{-1/3}$.
}
\label{chiphi:fig}
\end{figure}

The behavior of $\phi_i(t)$ is similar to that of a field
$q(x,t)$ describing a dynamical event in glassy systems (see 
Ref. \cite{ryamamoto}). In order to make the analogy more explicit, we 
consider the spatial average of $\phi_i$, $\bar\phi(t)=\sum_i 
\phi_i(t)/N$. In Fig.~\ref{phit:fig}, we show
the ensemble average $\bra \bar\phi(t) \ket $ for the system with 
$T = 10^{-a}$, where $a=3, 4, 5$, and $6$. The staying
time at the marginal saddle increases as $T$ is decreased.  
The graph of $\bra \bar \phi(t) \ket$ is similar to a time 
correlation function of density fluctuations in glassy systems.  
Furthermore, we measure the fluctuation intensity
\begin{equation}
\chi_ \phi(t)= N (\bra \bar \phi^2(t) \ket-\bra \bar \phi(t) \ket^2).
\end{equation}
As shown in  Fig.~\ref{chiphi:fig}, $\chi_\phi(t)$ has one peak 
at a time $t=t_*$, which 
provides a precise definition of $t_*$
intuitively  used in the previous paragraph. Then, $\chi_{\phi}(t_*)$  
increases as $T$ is decreased. In fact, the inset of 
Fig.~\ref{chiphi:fig}
suggests that $\chi_{\phi}(t_*)$ exhibits the power-law divergence
\begin{equation}
\chi_{\phi}(t_*) \simeq T^{-\gamma_\phi},
\label{gamma_phi}
\end{equation}
where the value of the exponent $\gamma_\phi$ is close to $1/3$. 
This divergent  behavior corresponds to the dynamical 
heterogeneity in glassy systems.  These observations naturally
lead us to a conjecture that the correlation length of $\phi_i(t_*)$
exhibits the power-law divergence as a function of $T$. 
In this paper, we do not investigate $\phi_i(t_*)$ directly but
propose a novel quantity that characterizes the divergent behavior.

\paragraph*{Result:}

Our basic idea for the characterization of the  
patterns is to employ
a special solution $\phi_*(t)$ of the equation $\partial_t \phi
=-\partial_\phi v(\phi)$ under the conditions $\phi(t) \to 1$ for
$t \to -\infty$, $\phi(t) \to 0$ for $t \to \infty$, and $\phi(0)=0.5$.
We then express the time evolution of the patterns as 
\begin{equation}
\phi_i(t)=\phi_*(t-\theta_i)+\rho_i(t-\theta_i),
\label{theta}
\end{equation}
where $\theta_i$ represents the exiting time from the marginal
saddle, which is defined by $\phi_i(\theta_i)=0.5$.
By this definition, the pattern of $\theta_i$ corresponds to 
the contour curve shown in Fig.~\ref{phase}.
For a typical trajectory whose weight is large, 
$\rho_i(t-\theta_i)$ is expected 
to be smaller than the special solution. 
Here, we conjecture that 
the fluctuation intensity of $\theta_i$ exhibits a power-law 
divergence of $T$
because the divergence is related to the Goldstone mode associated with the 
symmetry breaking of the  time translational symmetry.  
We thus characterize
the conjectured scale-free patterns  by the 
statistical quantities of $\theta_i$. 

\begin{figure}[htbp]
\includegraphics[width=7cm]{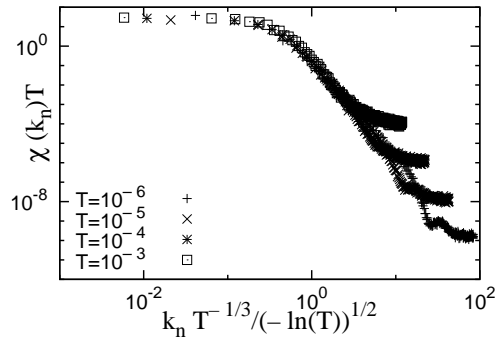}
\caption{
$ \chi(k_n;T) T$ as a function of $ k_n T^{-1/3}/(-\ln T)^{1/2}$ for
several values of $T$. 
$N=4096$. The statistical error-bar is about the symbol size.
By comparing the systems with $N=512, 
1024, 2048$, and $4096$,
it is found that the finite-size effects are fairly small.
}
\label{chi1d}
\end{figure}

Concretely, by introducing the Fourier transform of $\theta_i$ as
\begin{eqnarray}
\tilde \theta(k_n) 
& \equiv & \frac{1}{N}\sum_{j=1}^{N}  \e^{ik_n j}\theta_j,
\end{eqnarray}
we measure the spectrum  
\begin{eqnarray}
\chi(k_n;T)\equiv N \bra|\tilde \theta(k_n)|^2\ket,
\end{eqnarray}
where  $k_n  = 2 \pi n/N$ with $n=1,2,\cdots, N/2$.  
Here, $\bra  \ \ket$ represents the ensemble average.
In order to extract the singular behavior of $\chi(k_n;T)$  
 in the limit $T \to 0$, we first assume a scaling relation
\begin{eqnarray}
\chi(k_n;T)&=&T^{-\gamma}\tilde  \chi\lr{k_nT^{-\nu}},
\label{def_chi}
\end{eqnarray}
where the exponents $\gamma$ and $\nu$ characterize the divergences of 
the amplitude of the spectrum $\chi(k_n;T) $ and its length scale,
respectively,  in the limit $T \to 0$.
The exponents $\gamma$ and $\nu$ are determined such that  $\chi(k_n;T)$
with  $T=10^{-a}$, where $a=3, 4, 5$, and $6$, collapse into one universal 
curve. However, after some trials, we find that a logarithmic correction 
appears in the scaling relation.
To demonstrate it, 
we express  $\chi(k_n;T)T$ as a function of $k_n T^{-1/3}/(-\ln T)^{1/2}$
in Fig.~\ref{chi1d}, which suggests that 
the scaling relation (\ref{def_chi}) with the logarithmic correction
appears to be valid for the values
\begin{eqnarray}
\gamma &=& 1, 
\label{exp_ga} \\
\nu   &=& 1/3.
\label{exp_nu}
\end{eqnarray}


The exponent $\gamma$ is related to the scale dimension of 
the quantity $\theta_i$.  In order to confirm this fact 
explicitly, we consider the spatial correlation function 
\begin{eqnarray}
\bra \theta_l \theta_m\ket
&=& 
T^{-(\gamma-\nu)} \frac{\Delta K}{2\pi} \sum_{n=1}^{N} 
\e^{-iK_n (l-m) T^{\nu} } \bar \chi( K_n), 
\label{spec}
\end{eqnarray} 
where $K_n= k_n T^{-\nu}$ and $\Delta K= T^{-\nu} 2 \pi/N  $.
From  (\ref{spec}), we find a  
scaling form
\begin{equation}
\theta_l=T^{-(\gamma-\nu)/2}  \Theta(T^\nu l),
\label{scaling_theta}
\end{equation}
where $\Theta(x)$ is a fluctuating field whose 
distribution function is independent of $T$ in the limit 
$T \to 0$.  By combining  (\ref{theta})
with this scaling form (\ref{scaling_theta}), 
we arrive at our main claim
that the scale-free patterns  
are characterized by 
\begin{equation}
\phi_i(t) \simeq \phi_*\left (t-T^{-1/3}  \Theta(T^{1/3} i) \right),
\label{main}
\end{equation}
which clearly indicates the scale-free nature at 
 a saddle-node bifurcation point in the limit $T \to 0$.
We remark that the statistical quantities of $\phi_i$,
including the critical exponent $\gamma_{\phi}$ in  (\ref{gamma_phi}), 
can be  calculated theoretically from our result (\ref{main}). We  
believe that the representation (\ref{main}) captures the essence of 
the scale-free pattern.

\paragraph*{Concluding remarks:}


We have proposed a new universality class consisting of 
stochastic systems
undergoing a saddle-node bifurcation. 
A remarkable feature of this class is that the criticality 
near a saddle-node bifurcation originates from the fluctuations 
of the time passing through a marginal saddle.
We expect that there is a rich
variety of systems belonging to this class,
although a quantitative experimental measurement seems to 
be challenging.  An important
example includes neuronal avalanches \cite{Plentz}, 
which have recently been studied extensively. The analysis of 
a simple mean-field model  has revealed the critical nature 
clearly \cite{Ohta}. We conjecture that 
 scale-free patterns similar to those studied in this paper 
will be observed in systems related to neuronal avalanches.

Furthermore, we  have pointed out that the cooperative behavior 
observed in our model is related to the dynamical heterogeneity 
in glassy systems.  We wish to emphasize that 
this similarity is not superficial. In fact, recently, we have found
that the dynamics of $k$-core percolation in a random graph exhibits 
a jamming transition via a saddle-node bifurcation of some 
order parameter in the thermodynamic limit
\cite{kcore}. 

The most important result is the expression (\ref{main}), which claims 
that the criticality of the pattern observed at a particular 
time is characterized by $\theta_i$. 
The approach based on this idea  was  
addressed in the theoretical analysis of
a model \cite{epl}.  We expect that this approach would elucidate
the nature of the criticality of jamming systems.   


Before ending this paper, we present some important future problems.
The first problem is to understand the behavior of $d$-dimensional
systems. Theoretically, we should start estimating the upper-critical 
dimension by using physical arguments. The second problem is to uncover 
possible normal forms in multi-component systems. Even if the 
deterministic part exhibits a saddle-node bifurcation, the effect of 
noise is not determined uniquely in such systems. There might be
some systems in which the exponents are altered from the result
we obtained in this paper. 
Finally, the $\epsilon$-dependence of the length scale 
should be clarified. These will define another exponent, which might 
be more relevant in the context of critical phenomena. 

In order to consider these future problems, we need to develop the 
theoretical analysis. We have just finished the mean field analysis,
which will be published soon \cite{next}. According to the result, 
the field $\theta$ scales as $\theta \simeq T^{-1/3}$ and there are 
scaling relations as functions of $\epsilon T^{-2/3}$. From the naive 
dimensional analysis,
the length scale of the system with $\epsilon \not = 0$  turns out to be  
proportional to $\epsilon^{-1/2}$, which leads to $\gamma=1$ and $\nu=1/3$.  


This work was supported by a grant from 
the Ministry of Education, Science, Sports and Culture of Japan, 
No. 19540394. 
Mami Iwata acknowledges the support by Hayashi Memorial Foundation for 
Female Natural Scientists.



\begin{thebibliography}{99}
\bibitem{kappa}
However, according to our preliminary theoretical analysis, the
system with large $\kappa$ might exhibit different behavior,
which will be studied in future.
\end{thebibliography}
\end{document}